\documentclass[showpacs,aps,prd,epsf,showpacs,10pt]{revtex4}
\newcommand{\be}{\begin{equation}}
\newcommand{\ee}{\end{equation}}
\def\bq{\begin{eqnarray}}
\def\eq{\end{eqnarray}}
\def\beq{\begin{eqnarray}}
\def\eeq{\end{eqnarray}}
\def\ba{\begin{eqnarray}}
\def\ea{\end{eqnarray}}

\usepackage{dcolumn}
\usepackage{bm}
\newcommand{\s}{\ensuremath{\psi(t,r)}}
\newcommand{\n}{\ensuremath{\nu(t,r)}}

\newcommand{\dw}{\ensuremath{{d\Omega^2}}}

\begin{document}

\title{On trapped surface formation in gravitational collapse II}

\author{Pankaj S. Joshi}
\email{psj@tifr.res.in}

\affiliation{Tata Institute of Fundamental Research\\ Homi Bhabha Road,
Mumbai 400 005, India}

\author{Rituparno Goswami}
\email{rgoswami@phys.ualberta.ca}

\affiliation{Department of Physics\\ University of Alberta, Edmonton T6G 2G7, 
Canada}

\begin{abstract} Further to our consideration on trapped surfaces 
in gravitational collapse
~\cite{JG1}, 
where pressures were allowed to be negative while satisfying weak energy 
condition to avoid trapped surface formation, we discuss here several other 
attempts of similar nature in this direction. Certain astrophysical aspects 
are pointed out towards examining the physical realization of such a 
possibility in realistic gravitational collapse. 
\end{abstract}

\pacs{04.20.Dw, 04.70.-s, 04.70.Bw}
\maketitle

\section{Introduction}

We constructed recently a family of solutions to Einstein equations
which represented collapsing clouds with a perfect fluid form of
matter. While initially when the collapse commenced, the weak energy
condition, strong energy condition as well as the dominant energy 
conditions were all satisfied by the collapsing matter, subsequently during 
the evolution of collapse the pressure was allowed to turn negative, while 
still satisfying the weak energy condition. It was shown that the trapped 
surfaces and the formation of an event horizon are avoided in such 
a scenario.

One of the motivations here to consider negative pressure in 
late stages of gravitational collapse was the recent indication 
that the mini-superspace loop quantum gravity formalism implies generation 
of strong negative pressures, as induced by quantum gravity effects. 
This allows for the massive ejection of matter to the exterior spacetime 
in late stages of collapse, and a possible avoidance of the singularity 
in gravitational collapse
~\cite{JGP}. 
Singularity avoidance in cosmology has been examined earlier using 
a similar formalism
~\cite{Boj}.

Apart from this possibility that we explored in
~\cite{JG1}, 
there are several other mechanisms known, whereby the formation 
of trapped surfaces can be avoided in the context of spherically symmetric 
collapsing models, and which are solutions to Einstein equations. The 
purpose of this note is to discuss some of these to see how such mechanisms 
work dynamically, and towards examining the physical realizability of 
such a scenario in a physically realistic gravitational collapse.

We know that a general spherically symmetric comoving metric in 
four-dimensions can be written as,
\begin{equation}
ds^2=-e^{2\n}dt^2+e^{2\s}dr^2+R^2(t,r)\dw.
\label{eq:metric}
\end{equation}
Here $R(t,r)$ is the physical area radius while $r$ is the shell labelling 
comoving radial coordinate.
In general, the equation of apparent horizon in such a spherically 
symmetric spacetime is given as, 
\begin{equation}
g^{ik}\,R_{\, ,i}\,R_{\, ,k}=0.
\label{eq:hor}
\end{equation}
Thus we see that at the boundary of the trapped region 
the vector $R_{\, ,i}$ is null. Substituting (\ref{eq:metric}) 
in (\ref{eq:hor}) we get,
\begin{equation}
R'^{2}\,e^{-2\psi} - \dot{R}^{2}\,e^{-2\nu}\,=\,0
\label{eq:hor1}
\end{equation}
Now using Einstein equations we can write the 
equation of apparent horizon as
\begin{equation}
\frac{F}{R}=1.
\label{eq:apphorizon} 
\end{equation}
which gives the boundary of the trapped surface region of the 
spacetime. The function $F=F(t,r)$ in the above equation has an interpretation 
of the mass function for the cloud, which gives the total mass in 
a shell of comoving radius $r$, at an epoch $t$. The energy condition 
$\rho\ge0$ imply $F\ge0$ and $F'\ge0$. Since the area radius vanishes 
at the center of the cloud,
it is evident that in order to preserve the regularity of density and 
pressures 
at any non-singular epoch $t$, we must have  
$F(t,0)=0$, that is, the mass function should vanish at the center  
of the cloud.

Basically, the formation of trapped surfaces in spherically 
symmetric gravitational collapse can be viewed in terms of how much 
mass is there within a given area radius of the cloud, which decided 
whether a two-surface is trapped or not. Then, during the collapse 
if $F > R$, then the surface is trapped, otherwise it is not. 
It is thus clear that in the case of a continual collapse if trapped surfaces 
are not to develop, then we must have some mechanism available to throw 
away and radiate the mass so that trapping does not occur, and to 
preserve $R > F$ through out. Considering various works in this direction, 
such mechanisms can be basically divided into the following classes, 
mechanisms with non-constant comoving boundary and that with 
constant comoving boundary. As the names suggest these mechanisms differ 
considerably in their basic properties and also the dynamical 
evolution of interior matter is different. We give below a brief 
sketch of these mechanisms and discuss the physical pictures 
associated with them.

\section{Non-constant Comoving Boundary}

In a series of papers
~\cite{seno1}-~\cite{seno6} 
the authors here considered the boundary hypersurface of a collapsing star not 
co-moving with the interior matter. In that case it was proved that 
for a physical {\it timelike} matching hypersurface almost any interior 
spacetime obeying all energy conditions (including SEC) can be matched to 
an exterior 
outgoing Vaidya spacetime, which describes an outgoing unpolarized radiation, 
and the metric is given as,
\begin{equation}
ds^2_{+}=-\left(1-\frac{2M(V)}{r_v}\right)dV^2-2dVdr_v+r_v^2d\Omega^2
\label{eq:metricvaidya}
\end{equation} 
where $V$ is the retarded (exploding) null co-ordinate and $r_v$
is the Vaidya radius.

The authors used this property of matching hypersurface
to implicitly construct models of
continuous gravitational collapse, with complete evaporation of the
star, absence of closed trapped surfaces, and no singularity.
As described in these papers, 
one can always consider the boundary of a collapsing star 
as a general timelike hypersurface
not comoving with the interior matter, preserving the spherical symmetry. 
If we consider the intrinsic coordinates
of this hypersurface as $(\tau, \hat\theta, \hat\phi))$ then the
general parametric equations for this surface would be,
\begin{equation}
t=t(\tau);\;\; r=r(\tau);\;\; \theta=\hat\theta;\;\; \phi=\hat\phi.
\end{equation} 
This hypersurface is timelike whenever
\begin{equation}
\frac{\dot{r}}{\dot{t}}<e^{\nu-\psi},
\end{equation}
where the dot denotes differentiation with respect to the intrinsic time 
coordinate $\tau$. As an explicit illustration, one could consider the 
interior spacetime to be a collapsing Friedmann-Robertson-Walker (FRW) 
with the metric in the isotropic co-ordinates as,
\begin{equation} 
ds^2=-dt^2+\frac{a^2(t)}{b^2(r)}\left(dr^2+r^2\dw\right),
\end{equation}
where $b(r)=1+kr^2/4$. Then the required matching conditions would be,
\begin{equation}
(r_v)_\Sigma=\left(\frac{ar}{b}\right)_\Sigma
\end{equation} 
\begin{equation}
 M(V)_\Sigma=\left[\frac{ar^3(a_{,t}^2+k)}{2b^3}\right]_\Sigma
\end{equation}
\begin{equation}
(\dot{V})_\Sigma=-\left[\frac{a\dot{r}+b\dot{t}}{b^2(r/b)_{,r}-a_{,t}}\right]_\Sigma
\end{equation}
\begin{equation}
(\dot{r})_\Sigma=-\left[\frac{bp\dot{t}}{a\rho}\right]_\Sigma
\end{equation}
We can further impose an equation of state of the form $p=\gamma \rho$, 
and make the interior spacetime obey all the energy conditions.

Physically, for an observer co-moving with the interior matter, 
one may interpret that the boundary gradually evaporates away as the 
collapse proceeds and if the rate of evaporation
is greater than the rate of collapse then there need not be any
trapped surfaces forming and the star may completely evaporate without 
any singularity. Also there is an alternative interpretation for
an observer co-moving with the boundary hypersurface. Since the observer 
is not comoving with the interior matter now, the energy momentum tensor  
of the matter would not be diagonal, but generically of the form,
\begin{equation}
T^{ij}= (\rho+p)V^i V^j + pg^{i j} + \Phi(t,r)l^i l^j + \Pi^{ij}
\end{equation}
where $V^i$ is the timelike unit vector describing the motion of the 
hypersurface, 
$l^i$ is null vector and $\Pi^{ij}$ is the trace-free anisotropic
pressure. The observer would interpret this matter as a radiating general 
anisotropic fluid and the flux of energy across the matching hypersurface is
perfectly balanced so that the corresponding normal pressure (which
must be continuous due to Israel conditions) would be continuous.

As stated earlier, the mechanism here of throwing away matter  
depends only on the dynamics of the timelike matching hypersurface and not on 
the interior spacetime. In that case it is possible to find wide classes of 
collapsing interior spacetimes satisfying all energy conditions, which 
evaporate away completely without forming any trapped regions or singularity.
In such a case, the inner fluid becomes radiating and non-perfect if 
considered co-moving with the matching hypersurface by construction and 
the radiation coming from the interior crosses the surface of the star 
and escapes away. It should be noted that in this interpretation the inner 
fluid is, however, not comoving with observers with 4-velocity $v=d/dt$, 
at almost all points in the interior, and the dynamics of the matching 
hypersurface and the rate of outgoing radiation depend on the choice of 
interior solution.

\section{Bulk Mechanisms}

Another known mechanism to avoid trapped surface formation is to include 
radial heat flow in the interior, again satisfying all the energy conditions. 
This radial heat flux is generated by various dissipative processes in the 
collapsing interior (for example, neutrino radiative heat conduction), and 
which may play a dramatic role in the final outcome of the collapse. 
For example, the solution presented in 
~\cite{Las} 
describes the collapsing matter which radiates out mass due to the 
outward heat flow to an exterior Vaidya region 
and through this process the complete star evaporates away leaving behind a 
flat Minkowski spacetime. 
A different solution was found in 
~\cite{ban1} and~\cite{ban2}, 
where the loss of matter due to heat flux prevents  
the trapped surface formation and a naked singularity is formed as the 
end state of the gravitational collapse. 
Detailed investigations on the dissipative processes in the 
interior spacetime, 
using diffusion approximation was carried out in 
~\cite{herr1}-~\cite{Herr5},
and it was shown that due to the {\it inertia of heat} (as described by Tolman 
~\cite{tol}), 
the collapse may slow down and stop before the singularity formation.

In presence of the radial heat flow, the $T_{tr}$ component 
of the Energy-Momentum tensor is non-zero necessarily. 
For example if we consider a collapsing general anisotropic fluid with 
heat flux, 
the Energy-Momentum tensor is given as
\begin{equation}
T^{ik}=(\rho + p)V^iV^k+pg^{ik}+q^iV^k+V^iq^k + \Pi^{ij},
\label{eq:em}
\end{equation} 
where $V^i$ is the timelike unit vector, $q^i$ is the heat flux which 
can be always
written as (by redefining the density and pressure),
\begin{equation}
q^i=Q(t,r)\delta^i_r
\label{eq:q}
\end{equation}
In this case, we can match the collapsing interior with Vaidya spacetime as 
the exterior at the boundary hypersurface $\Sigma$ given by $r=r_b$.
We note that now we can make the matching hypersurface co-moving with the 
matter, 
that is, the co-moving boundary of the cloud remains fixed. 
For the required matching we use the {\it Israel-Darmois} conditions, 
where for the interior and exterior spacetimes, the first and second 
fundamental forms are matched (which are the metric coefficients and 
the extrinsic curvature respectively), 
at the boundary of the cloud. As there is a flux of unpolarized 
radiation going radially 
outwards from each point of the cloud, the exterior would be an 
outgoing Vaidya spacetime.

As an explicit illustration, if we consider the interior spacetime to be 
shearfree with a general metric (see ~\cite{Herr5}, ~\cite{RG} for 
more general 
spacetimes),
\begin{equation}
ds^2=-e^{2\n}dt^2+e^{2\s}\left[dr^2+r^2\dw\right],
\label{eq:metric1}
\end{equation}
the matching conditions are given by 
(see ~\cite{San} for details),
\begin{eqnarray}
[Qe^\psi]_\Sigma=(p)_\Sigma ; & R_\Sigma=(r_v)_\Sigma ; & F_\Sigma=2M(V)_\Sigma
\label{eq:match1}
\end{eqnarray} 
\begin{equation}
[e^\nu]_\Sigma dt=\left(1-\frac{2M(V)}{r_v}+2\frac{dr_v}{dV}\right)_\Sigma dV.
\label{eq:match2}
\end{equation} 
Thus we see that the matter is radiated away in such dissipative 
mechanisms and the trapping can be avoided till the singularity, if there 
is one. Also, the interior spacetime would then satisfy all the energy 
conditions (both SEC and DEC). 
>From equation (\ref{eq:match1}) it is clear that 
if we have $Q=0$ and the co-moving boundary of the cloud is fixed, 
then we have $p=0$ at boundary, which implies that one does
not have any radiation in the Vaidya regeion, or in other words the 
exterior necessarily becomes Schwarzschild.

The general properties of collapse in the presence of heat flux 
can be very different from when it is absent. Heat flux can act as a
source of spacetime shear and also as a {\it frictional} term, which may 
slow down or stop the collapse 
~\cite{Herr5}, ~\cite{RG}. 
If we couple the Einstein equations with {\it Israel-Stewart}
transport equation 
~\cite{israel} 
(assuming the diffusion approximation)
one can easily see that the {\it effective inertial 
mass density} of the infalling matter (as defined by the `Newtonian' form of 
the equation of motion) reduces due to the presence of heat flow and may
become negative as the collapse proceeds. For a perfect fluid, 
the effective inertial mass density behaves as,
\begin{equation}
(\rho+p)_{eff}=(\rho+p)(1-\alpha);\; \alpha=\frac{\kappa T}{\tau (\rho+p)}
\end{equation}
where, $T(t,r)$ is the temperature, $\tau$ is the relaxation time and 
$\kappa$ is the thermal conductivity.

As pointed out by these authors, if we consider adiabatic collapse, 
the conversion of the binding energy of the star to the internal energy, 
would lead to increase of the temperature enormously, which in turn 
can make $\alpha>1$. At such high temperatures the matter can actually be 
radiated away at an enormous rate and may stop the trapped surface 
formation.

\section{Negative pressure in the interior}

Finally, in a kind of bulk mechanism to radiate away the matter
as the collapse progresses, one can allow the pressure of the interior
spacetime to have a negative value, which would imply from the
Einstein equations,  
\be
\dot{F} < 0,
\ee
and the interior will continuously throw away the matter, without
having recourse to or need to have heat flux. However, in such a 
case, the matter need not obey the strong energy condition, but   
could obey the weak energy condition and the dominant energy condition. 
In that case, the comoving boundary surface can be kept at a constant
radius, $r_b=const.$ as in the first case, and obviously without 
any heat flux we cannot match the spacetime to a Vaidya exterior. However, 
the matching of the spacetime in this situation would be with a generalized 
Vaidya metric (which describes a large class of spacetimes with  
matter which is a combination of matter fields of {\it type I} and 
{\it type II}), as we have discussed in~\cite{JG1}.

The idea here is not to construct yet another class of solutions to 
Einstein equations, where trapped surfaces do not form in dynamical collapse
evolution. One would rather want to invoke a physical mechanism, such
as negative pressures as induced through the incorporation of quantum effects,
towards achieving the avoidance of trapped surfaces formation in
a gravitational collapse. 
In this case also, the negative pressure acts as a frictional term 
which decreases the rate of collapse and there may be a `rebounce'. Such a 
rebounce was discussed in
~\cite{GP} 
where the negative pressure due to loop quantum gravity effects 
stops the collapse and also the singularity formation.

\section{Discussion}

We note that there have also been some attempts recently, to remove the 
horizons non-classically, invoking some kind of a quantum or string theoretic 
framework. For example, in
~\cite{Math}, 
a `fuzz-ball' kind of a model tries to replace the horizon and its
interior, towards trying to resolve the information paradox.

We like to note here an astrophysical aspect related to physically 
realistic gravitational collapse of massive stars, and its relevance to some 
of the models above. While all the mechanisms such as the ones discussed 
above, are valid as theoretical constructs which are solutions to the 
Einstein equations, from an astrophysical perspective, what happens is 
the following. A massive star, which one has tried to model through 
some of these mechanisms, starts a continual collapse at the end of its life 
cycle. While during its regular life it was radiating, which can again be 
modelled possibly by a suitable Vaidya exterior, the total amount of mass 
loss will be rather small during such a phase, as compared to the total 
mass of the star. Hence, the amount of mass that must be radiated away, 
in order for the trapped surfaces formation to be actually avoided after
the final collapse starts has got to be rather large, and an extremely 
efficient mass radiation must take place. There is another important 
constraint also. The entire mass ejection such as above must take 
place within the matter of a few seconds, or at best tens of seconds, 
which will be the duration of the final catastrophic collapse. The massive 
star which has lived millions of years will now collapse towards the end 
of its life, in a matter of order of seconds. Hence, tens of solar masses, 
for a massive enough star, must be radiated away in few seconds. No 
astrophysically viable mechanisms are still envisaged or likely to exist
at a purely classical level to achieve such an enormous mass loss in 
such a short time scale.

Therefore, it is quite possible that while radiation models 
such as above are interesting as theoretical constructs, none of these could 
be in fact taken sufficiently seriously, so as to enable the avoidance of 
trapped surfaces altogether in the final stages of a realistic gravitational 
collapse, and these 
may not be able to delay or avoid trapped surfaces in reality. That is, 
when a massive star starts collapsing, trapped surfaces must develop, despite 
some radiation being thrown out, because too much of matter is accumulating 
and accrediting too fast and in too short a time, and not enough can be 
radiated away only through the mechanisms such as those discussed above 
so as to avoid or delay trapping this way.
In our view, this has been one of the main motivations for the black hole
physics, namely that trapped surfaces formation cannot be
avoided due to very short time scales of catastrophic gravitational
collapse of a massive star towards the end of its life cycle.

An important alternative possibility in this direction is
that, during the collapse the trapped surfaces formation can be actually 
delayed, depending on the distortion of the trapped surface geometry as 
caused by inhomogeneity in the matter distribution and the effects such as 
spacetime shear
~\cite{JDR}. 
While pure radiation mechanisms such as above may not be able 
to achieve enough mass emission, the shear or inhomogeneity can independently 
distort the trapped surface geometry as much as to make the final 
singularity of collapse visible at the classical level, rather than the 
trapped surfaces or event horizon covering it fully. To give an explicit 
example, it is seen in dust collapse that while there is no radiation to 
the exterior which is a Schwarzschild geometry, the inhomogeneity in the 
density distribution by itself delays the trapped surfaces, thus 
allowing the singularity to be visible.

It is for this reason that, some kind of a quantum gravity mechanism
has to be invoked in the very late stages of collapse towards creating the 
required mass loss from the regions near the visible singularity, or 
else the trapped surfaces must form at some stage in collapse. These radiation 
mechanisms can, however, help along with shear or inhomogeneity in delaying
trapping because they cause at least some mass loss. This can add
to the delaying of trapped surface formation so that the quantum effects
in the ultra-strong gravity regions may be actually visible. Therefore,
these theoretical possibilities have to be looked into carefully.

Acknowledgement: It is a pleasure to thank F.Fayos, L. Herrera, 
R. T. Herrera, and J. M. M. Senovilla for many discussions and comments.

\end{document}